\documentclass[a4paper]{spie}  

\usepackage{amsmath,amsfonts,amssymb}
\usepackage{graphicx}
\usepackage[colorlinks=true, allcolors=blue]{hyperref}
\usepackage{subfig}
\usepackage[font=footnotesize]{caption}

\def\5{\footnotesize V\normalsize}
\def\4{\footnotesize IV\normalsize}
\def\3{\footnotesize III\normalsize}
\def\2{\footnotesize II\normalsize}
\def\1{\footnotesize I\normalsize}

\makeatletter
\renewcommand{\maketitle}{\bgroup\setlength{\parindent}{12pt}
\begin{center}
\textbf{\@title}
\end{center}
\begin{flushleft}
\@author
\end{flushleft}\egroup
}
\makeatother

\title{Stellar astrophysics in the near UV with VLT-CUBES}
\author{{H. Ernandes$^{1,2,3}$, C. J. Evans$^{2}$, B. Barbuy$^{1}$, B. Castilho$^{4}$, G. Cescutti$^{5}$, 
N. Christlieb$^{6}$, S. Cristiani$^{5}$, G. Cupani$^{5}$, P.
Di Marcantonio$^{5}$, M. Franchini$^{5}$,C. Hansen$^{7}$,
A. Quirrenbach$^{6}$, R. Smiljanic$^{8}$ } \\
{\footnotesize $^{1}$Universidade de S\~ao Paulo, IAG, Rua do Mat\~ao 1226, Cidade Universit\'aria, S\~ao Paulo, 05508-900, Brazil\\
$^{2}$UK Astronomy Technology Centre, Royal Observatory, Blackford Hill, Edinburgh, EH9 3HJ, UK\\
$^{3}$IfA, University of Edinburgh, Royal Observatory, Blackford Hill, Edinburgh, EH9 3HJ, UK\\
$^{4}$Laborat\'orio Nacional de Astrof\'isica/MCTIC, Rua Estados Unidos, 154 - 37504-364, Itajub\'a, MG, Brazil\\
$^{5}$INAF - Osservatorio Astronomico di Trieste, via G. B. Tiepolo 11, 34131 Trieste, Italy\\
$^{6}$Landessternwarte, Zentrum für Astronomie der Universit\"at Heidelberg, K\"oniggstuhl 12, 69117, Heidelberg, Germany\\
$^{7}$Max Planck Institute for Astronomy,Koenigstuhl 17, 69117 Heidelberg Germany\\
$^{8}$Nicolaus Copernicus Astronomical Center, Polish Academy of Sciences, Bartycka 18, 00-716, Warsaw, Poland \\
}
}

\begin{document} 

\maketitle

\begin{abstract}
Alongside future observations with the new European Extremely Large Telescope (ELT), optimised instruments on the 8-10m generation of telescopes will still be competitive at `ground UV' wavelengths (3000-4000\,\AA). The near UV provides a wealth of unique information on the nucleosynthesis of iron-peak elements, molecules, and neutron-capture elements. In the context of development of the near-UV CUBES spectrograph for ESO's Very Large Telescope (VLT), we are investigating the impact of spectral resolution on the ability to estimate chemical abundances for beryllium and more than 30 iron-peak and heavy elements. From work ahead of the Phase~A conceptual design of CUBES, here we present a comparison of the elements observable at the notional resolving power of CUBES ($R$\,$\sim$\,20,000) to those with VLT-UVES ($R$\,$\sim$\,40,000). For most of the considered lines signal-to-noise is a more critical factor than resolution. We summarise the elements accessible with CUBES, several of which (e.g. Be, Ge, Hf) are now the focus of 
quantitative simulations as part of the ongoing Phase A study.
\end{abstract}


\section{INTRODUCTION}
\label{sec:intro}  

The 2020s will see the first operations of the European Extremely Large Telescope (ELT), starting a significant new era for observational astronomy. With a primary aperture of 39\,m and adaptive optics to correct for atmospheric turbulence, the ELT will provide an unprecedented combination of sensitivity and exquisite angular resolution. However, to deliver good performance across a wide wavelength range, four of the five ELT mirrors will be coated with  protected silver, resulting in diminished performance at shorter wavelengths ($<$4500\,\AA) compared to aluminium-coated mirrors. As a result, an optimised spectrograph on the VLT can be competitive with the ELT for observations shortwards of $\sim$4000\,\AA\ (e.g. Evans et al. 2016).

We are mid-way through a Phase A conceptual design study of the Cassegrain U-Band Efficient Spectrograph (CUBES) for the VLT. This builds on a previous Phase A study of a near-UV spectrograph for the VLT (Barbuy et al. 2014), with an updated overview of the scientific motivations given by Evans et al. (2018). The key aspect for stellar astrophysics is that the near UV provides us with valuable information on iron-peak and heavy elements, as well as some lighter elements (notably Beryllium) and light-element molecules (CO, CN, OH). Indeed, some of the neutron-capture elements only have lines in the near UV (see, e.g. Sneden et al. 2003).

Advanced simulations to finalise the instrument requirements and quantify the performance of CUBES are now underway as part of the Phase A study. Here we present analysis that provided the starting
point of the conceptual design. The over-arching design goal for CUBES is to maximise the instrument efficiency in the near UV, while still enabling quantitative stellar abundances, so spectral resolving power ($R$) is a key parameter. One of the leading instruments for stellar abundances at such wavelengths has been the Ultraviolet and Visual Echelle Spectrograph (UVES) on the VLT (Dekker et al. 2000), typically providing $R$\,$=$\,40\,000 with a 1$''$ slit. Abundances for many elements can potentially be estimated at lower resolution, but we need to investigate their feasibility on a line-by-line basis. Here we present a first study of the impact of spectral resolution on a broad range of elemental diagnostic lines in the near UV (over the range 3020-4000\,\AA). Specifically, we investigate which are accessible with $R$\,$=$\,20\,000 from CUBES compared to $R$\,$=$\,40\,000 with UVES.

\section{Stellar nucleosynthesis}

Understanding the origins of the elements in the periodic table has a prominent role in  both astronomy and nuclear physics. Indeed, production of the chemical elements that we are made of and use every day in our lives is one of the most profound questions we can ask. Each isotope has a complex production channel, which includes numerous nuclear reactions. 

\begin{figure}[!h]
    \centering
    \includegraphics[trim= 0 0 0 1.75cm, clip, width=150mm]{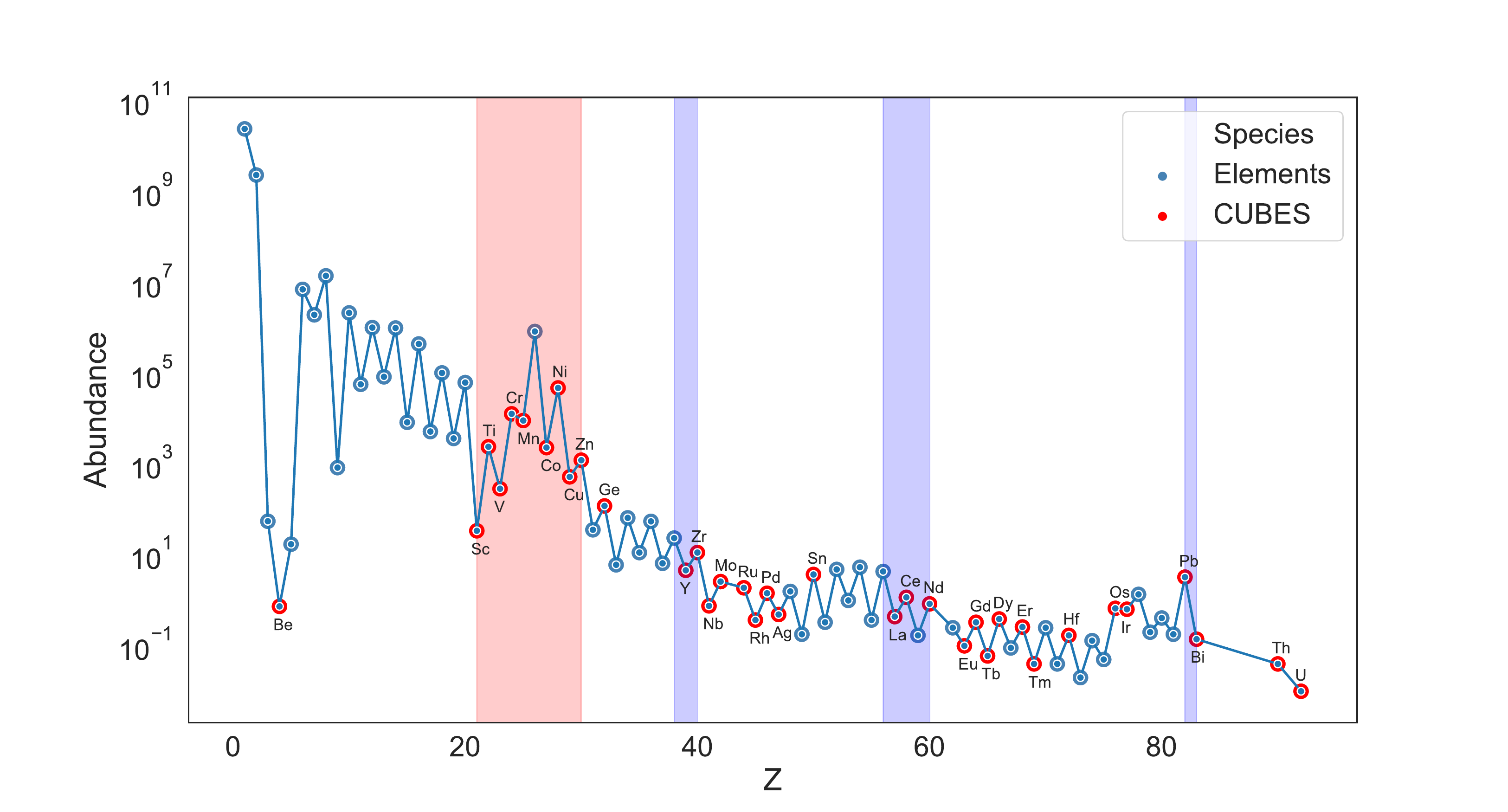}
    \caption{Abundances of elements in the Solar System vs. atomic number (Z) normalised by the abundance of $^{28}$Si to 10$^{6}$ (Lodders, 2003). Iron-peak elements and the three peaks of the s-process are highlighted by the pink and blue panels, respectively. Elements with near-UV (3020-4000\,\AA) spectral lines and observable with CUBES (at $R$\,$=$\,20\,000) are indicated in red.} 
    \label{rpeaksZ}
\end{figure}

To simplify the problem somewhat, we can group elements together that have common formation mechanisms. For instance, the lightest elements (H, He, Li) were formed in the first minutes after the Big Bang (with some debate remaining as to whether trace amounts of Be are also primordial). He is also produced via the proton-proton chain in main-sequence stars, whereas Li and Be are produced and destroyed by this process, which depletes them even further than their otherwise low relative abundances (see Fig.~\ref{rpeaksZ}). The light $\alpha$-elements typically have spectral lines in the visible and infrared, but elements around the iron-peak (21\,$\leq$\,Z\,$\leq$\,30, highlighted in pink in Fig.~\ref{rpeaksZ}, in which the most abundant element is Fe itself), also have absorption lines in the near-UV regime.

Beyond the iron peak many elements only have detectable lines at ground-UV wavelengths and into the space UV (e.g. Sneden et al. 2003). Fusion reactions beyond the iron peak (i.e. Z\,$>$\,30) are endothermic and would also have to overcome the Coulomb barrier, so these elements are generally not formed by proton capture. Production of such elements therefore occurs via neutron-capture nucleosynthesis, which is described by two major mechanisms, the rapid and slow capture of neutrons (r-process and s-process, respectively). The s-process occurs when the neutron-capture timescale is much lower than that for $\beta$-decay ($\tau_{n}\,\gg\,\tau_{\beta}$), hence this process flows in the the valley of beta stability. The three peaks of the s-process (highlighted by the blue panels in Fig.~\ref{rpeaksZ}) appear due to the bottleneck effect of the magic numbers 50, 82 and 126. The r-process is defined by the converse, $\tau_{n}\,\ll\,\tau_{\beta}$, where neutron capture occurs before
nuclei have time to undergo $\beta$-decay. Given these timescales, the two processes are associated with very different astrophysical environments.

To address fundamental questions such as the origins of the heavy elements and their complex nucleosynthesis we need access to the wealth of information that near-UV spectra contain. For example, the chemical abundances of metal-poor stars provide us with valuable probes of the nucleosynthesis processes of the first stars and the early evolution of the Milky Way, but near-UV observations are limited to only small samples (tens of stars) with current instrumentation.
Ahead of investigating the required spectral resolution for near-UV spectroscopy and to identify key diagnostic lines, we briefly outline some of the cases motivating new observations.

\subsection{Beryllium}
Although one of the lightest, simplest elements, there remain profound questions regarding the production of Beryllium in the early Universe. For instance, the recent upper limit for the Be abundance in an extremely metal-poor star ([Fe/H]\,$=$\,$-$3.84) from Spite et al. (2019) is consistent with no primordial production, but larger samples of very metal-poor stars are required to constrain its formation channels. Moreover, Be is a potentially powerful tracer of Galactic and stellar physics in a range of different contexts, including stellar evolution, the formation of globular clusters,  and the star-formation rate and chemical evolution of the Galaxy (Smiljanic 2014).

The only Be lines available from the ground for abundance estimates are two Be~\2 resonance lines at 3130.42, 3131.06\,\AA, which require good S/N ($\gtrsim$50) and sufficient resolution to clearly discern them from nearby, relatively strong V~\2 ($\lambda$3130.3) and Ti~\2 ($\lambda$3130.8) absorption. There are only $\sim$200 stars with estimated Be abundances (from Keck-HIRES and VLT-UVES), which span near-solar metallicities down to [Fe/H]\,$<$\,$-$3 (Boesgaard et al. 1999, 2011; Primas et al. 2000a,b; Smiljanic et al. 2009). The limiting magnitude of current observations is $V$\,$\sim$12\,mag; observations down to at least three magnitudes deeper with CUBES will provide the large homogeneous samples required, particularly at the metal-poor end given discoveries from ongoing wide-field, multi-band photometric surveys (e.g. Pristine, Starkenburg et al. 2017;  SkyMapper, Da Costa et al. 2019; Wolf et al. 2018).


\subsection{Iron-peak}
The iron-peak group is divided in two: the lower group (21\,$\leq$\,Z\,$\leq$26), which is mainly produced by explosive oxygen and silicon burning (Nomoto, Kobayashi \& Tominaga, 2013) and the upper group (27\,$\leq$\,Z\,$\leq$\,32) which are synthesized in two processes: $\alpha$-rich freeze-out and the weak s-process (Woosley, Heger \& Weaver, 2002; Limongi \& Chieffi, 2003); see also the review by Barbuy et al. (2018) and recent results for Sc, V. Mn, Cu, and Zn in globular clusters in the Galactic bulge (Ernandes et al. 2018). The near UV is less critical for these elements, but can still provide useful information for species such as Zn\,\1.

\subsection{Heavy elements}\label{heavy}
As previously mentioned, heavy elements (i.e. Z\,$>$\,30) are produced by two
major mechanisms, the r-process and s-process. The s-process is typically associated with stars on the Asymptotic Giant Branch (AGB, e.g. Busso, Gallino \& Wasserburg, 1999) and includes the elements highlighted by the blue panels in Fig.~\ref{rpeaksZ}. The s-process also occurs in massive stars during their He-burning phase. They mainly produce first-peak s-process elements (Sr, Y and Zr), and have an impact at low metallicity thanks to rapid rotation (Frischknecht et al. 2012, Limongi \& Chieffi 2018).

The formation channels for the r-process are particularly topical given the detection of the GW170817 kilonova from a binary neutron-star merger (Pian et al. 2017; Smartt et al. 2017; Watson et al 2019). The r-process is thought to occur both during the merging and in the milliseconds afterwards (e.g. Bovard et al. 2017), and is thought to play an important role in the chemical evolution of the Galaxy (Matteucci et al. 2014; Cescutti et al. 2015). Other predicted sites of r-process nucleosynthesis include magnetohydrodynamically-driven jets from core-collapse SNe, resulting from rapidly-rotating massive stars with a strong magnetic field (Winteler et al. 2012; Nishimura, Takiwaki \& Thielemann, 2015) and accretion discs in the supernova-triggering collapse of rapidly-rotating massive stars (or collapsars, Siegel et al. 2019).

\subsection{CNO abundances}

\begin{figure}[h]
    \centering
    \includegraphics[trim=0 0 0 1.75cm, clip, width=0.7\columnwidth]{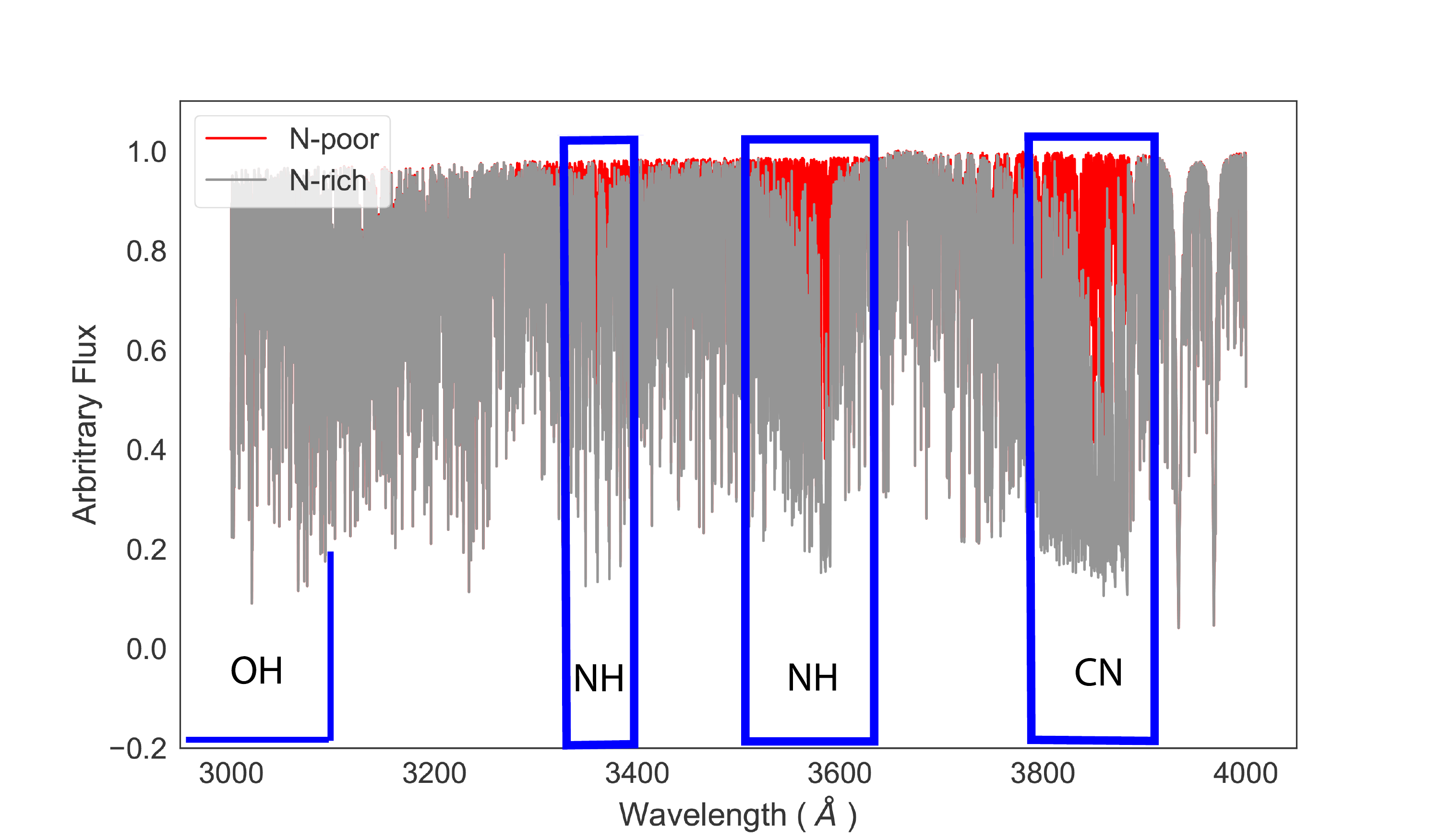}
    \caption{{\it Red:} Synthetic spectrum of a metal-poor star generated using the 
    {\sc turbospectrum} radiative transfer code (Plez, 2012), adopting physical  parameters as for CS\,31082-001 (Cayrel et al. 2001; Hill et al. 2002) including [Fe/H]\,$=$\,$-$2.9. {\it Grey:} Synthetic spectrum of the same star, but now N-rich ($\Delta$[N/H]\,$=$\,$+$2.0\,dex).}\label{CNO}
\end{figure}

Abundances of CNO bring a wealth of information on stellar evolution and the chemical evolution of the Galaxy. In contrast to the atomic transitions of the elements discussed above, CNO features in the near UV are dominated by a series of molecular bands (see Fig.~\ref{CNO}), which include the A-X OH transitions at the shortest wavelengths can be used to estimate oxygen abundances.

Many potential targets in this context will be drawn from the so-called
Carbon-enhanced metal-poor (CEMP) stars, which have [C/Fe]\,$>$\,$+$1.0 (see Beers \& Christlieb, 2005). Although rare, they demonstrate a diverse range of abundances of neutron-capture elements, commonly grouped as: `CEMP-no' (no over-abundance of r-process elements), `CEMP-r' and `CEMP-s' (stars with over-abundances of r- and s-processed elements, respectively) and `CEMP-r/s' (with apparent contributions from both processes enriching their photospheres). 

A range of scenarios have been explored to investigate these patterns, including rotational mixing in rapidly rotating, low-metallicity stars (e.g. Chiappini, 2013; Choplin et al. 2016) and supernova models which include both mixing and fallback of material to yield the observed abundance ratios (e.g. Umeda \& Nomoto, 2002, 2005; Tominaga, Iwamoto \& Nomoto, 2014). In short, the CEMP stars are perfect probes to investigate nucleosynthesis from the first stars (including production of neutron-capture elements) as well as mass transfer in binary systems (e.g. Abate et al. 2015). However, comprehensive near-UV spectroscopy of CEMP stars to date has been limited by the sensitivity of current facilities to a few relatively bright targets (e.g. Placco et al. 2015; T. Hansen et al. 2015; Hansen et al. 2019).

\section{Spectral Analysis}

\subsection{Simulated spectra}

To investigate the feasibility of abundance estimates for different elements as a function of spectral resolving power, we created small grids of synthetic spectra with the {\sc pfant} code, using model atmospheres interpolated from the grid of 1D, hydrostatic, LTE {\sc marcs} models from Gustafsson et al. (2008). Atomic data for the calculations were taken from the VALD database (Ryabchikova et al. 2015). We calculated two sets of spectra that will be illustrative of CUBES observations, with effective temperatures (T$_{\rm eff}$) and surface gravities (log\,$g$) appropriate for a G-type dwarf and a K-type giant, with two metallicities (as traced by the iron abundance, [Fe/H]), as summarised in Table~\ref{models}. A microturbulence ($v_{\rm turb}$) of 2.0\,km\,s$^{-1}$ was adopted in all  calculations.

Relative to the solar-scaled abundances (defined by [Fe/H]), the abundances for a broad range of elements with near-UV absorption lines were varied to investigate the feasibility of observations (and responsiveness of the lines to abundance changes). We calculated models for both the dwarf and giant templates, at both metallicities and spectral resolving powers, varying the abundances of 39 elements\footnote{Be, Sc, Ti, V, Cr, Mn, Co, Ni. Cu, Zn, Ge, Y, Zr, Nb, Mo, Ru, Rh, Pd, Ag, Sn, Ba, La, Nd, Sm, Eu, Gd, Tb, Dy, Ho, Er, Tm, Yb, Hf, Os, Ir, Pb, Bi, Th, and U.} simultaneously by $-$0.5, 0.0, $+$0.5, and $+$1.0\,dex. The full list of elements and absorption lines considered are detailed in Table~\ref{ablines} in the Appendix.

\begin{figure}[h]
    \centering
    \includegraphics[trim= 0 0 0 0.66cm, clip,width=150mm]{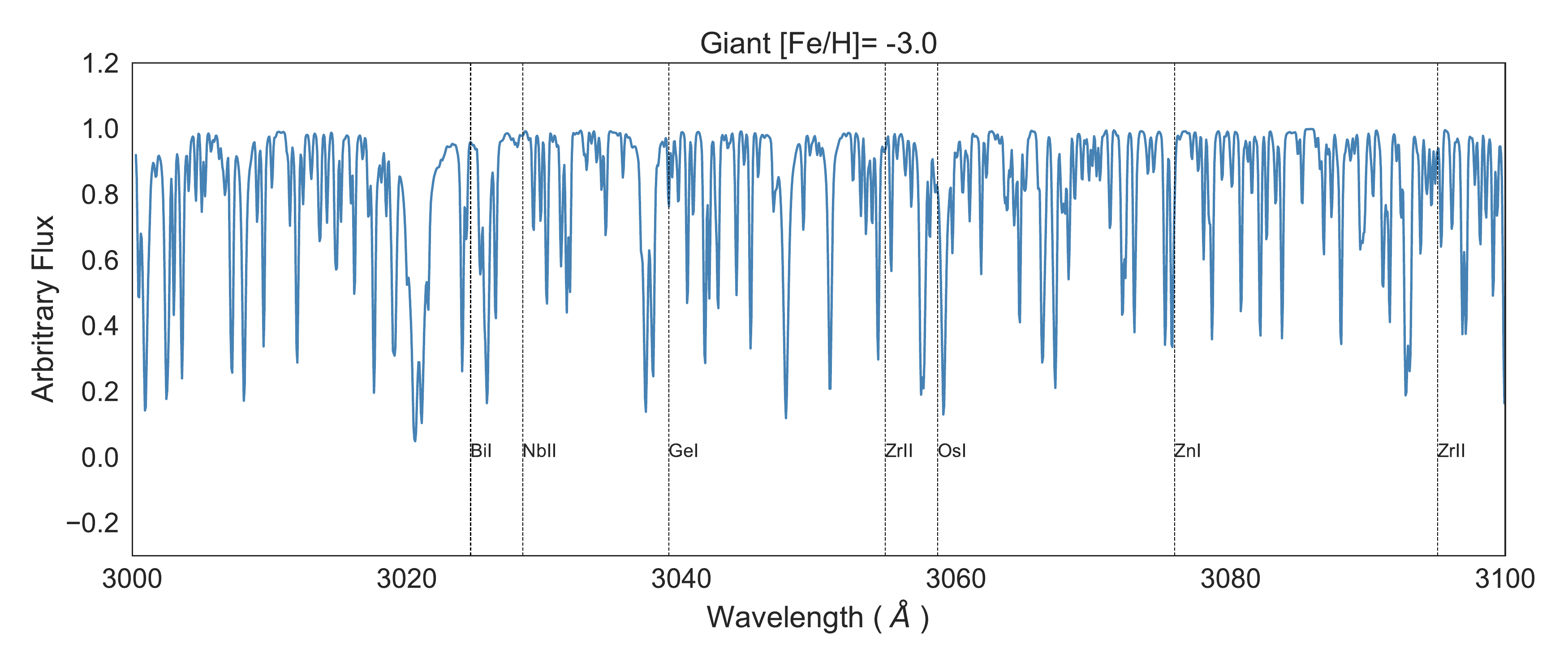}
    \caption{Example section of metal-poor ([Fe/H]\,$=$\,$-$3.0) giant spectrum at R\,$=$\,20\,000, prior to introduction of noise.}\label{allspec1}
\end{figure}

\begin{figure*}
    \centering
    \subfloat[Ge\,{\scriptsize I} $\lambda$3039.07]{
  \includegraphics[width=0.25\columnwidth]{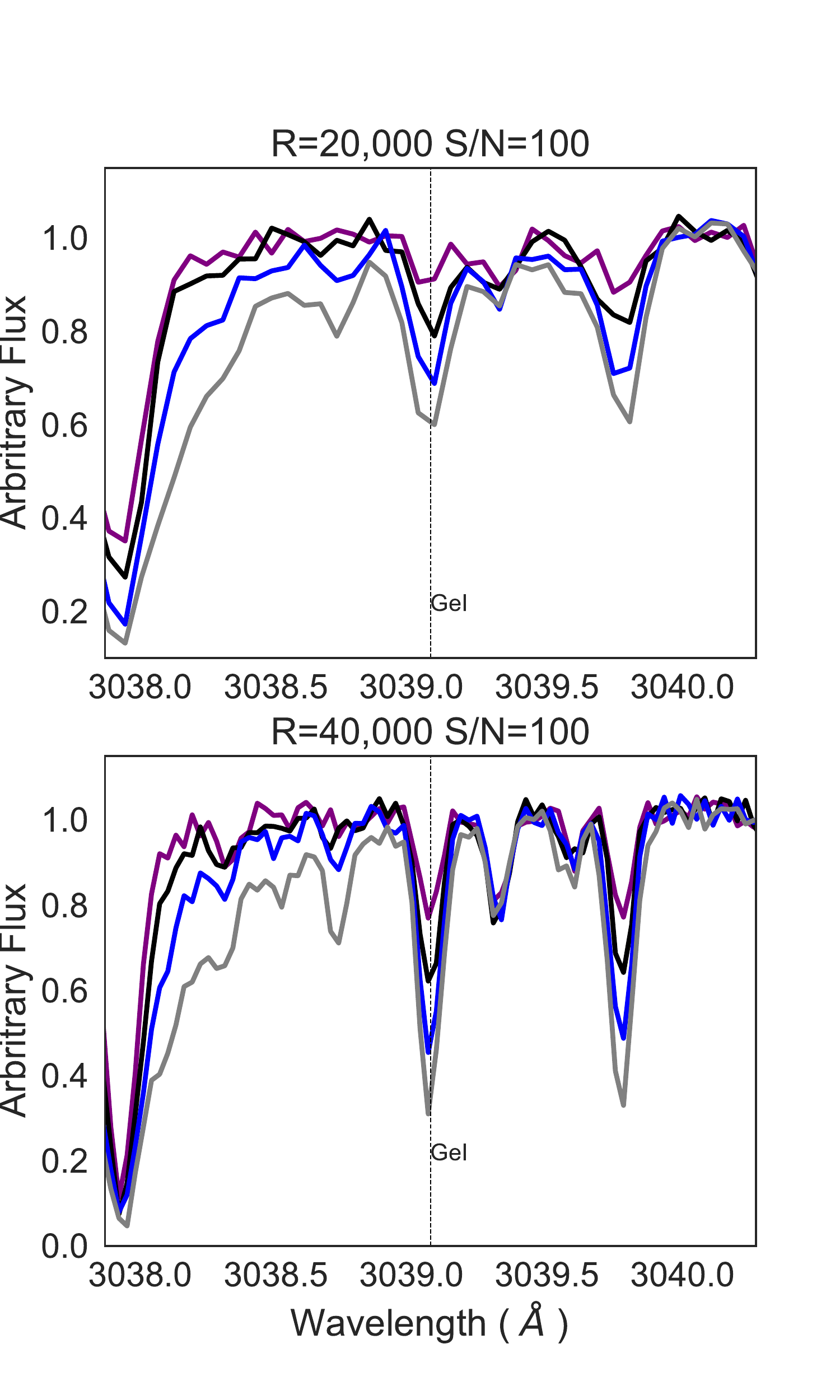}}
\subfloat[Co\,{\scriptsize I} $\lambda$3529.03]{
  \includegraphics[width=0.25\columnwidth]{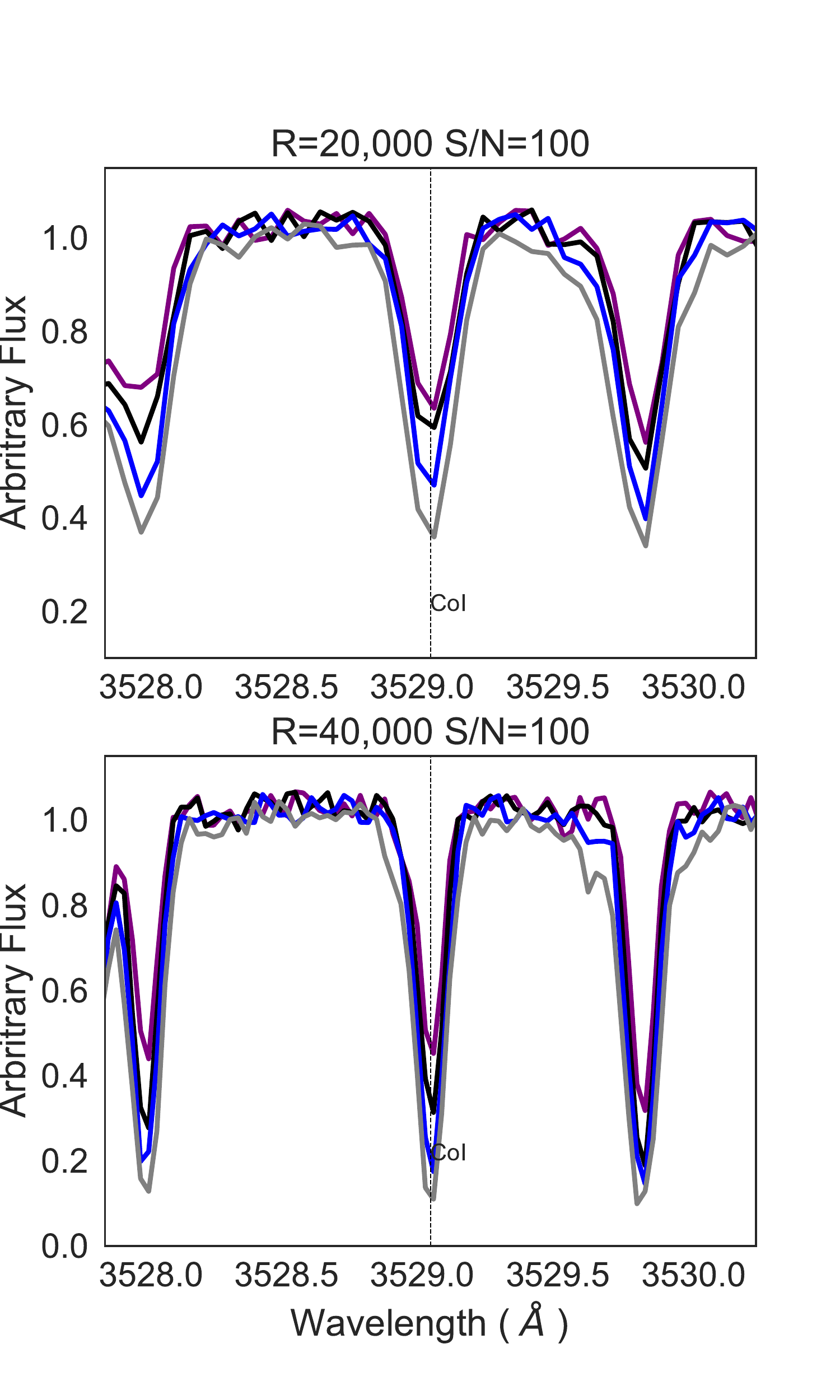}}
\subfloat[Ni\,{\scriptsize I} $\lambda$3597.71]{
  \includegraphics[width=0.25\columnwidth]{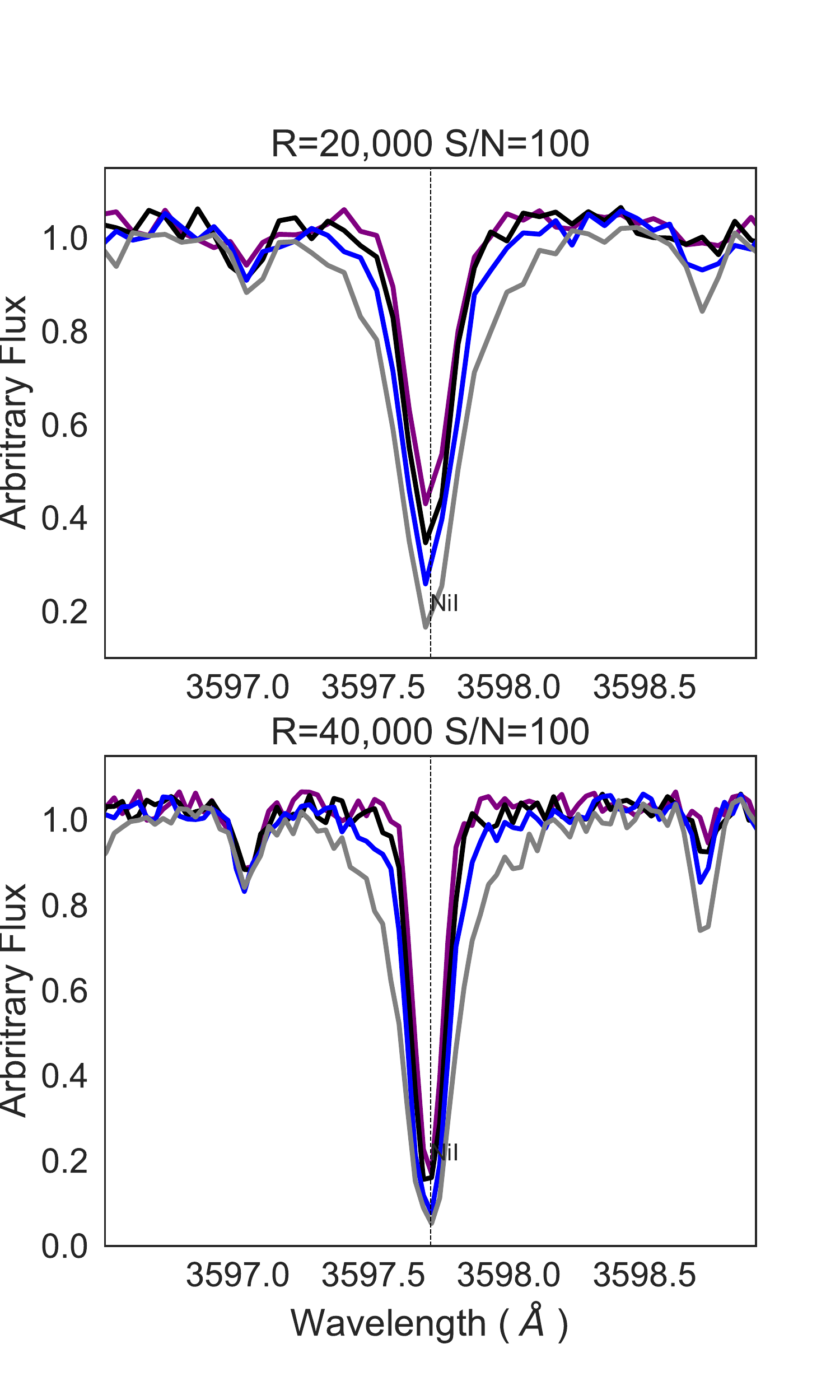}}
\subfloat[Y\,{\scriptsize II} $\lambda$3600.74]{
  \includegraphics[width=0.25\columnwidth]{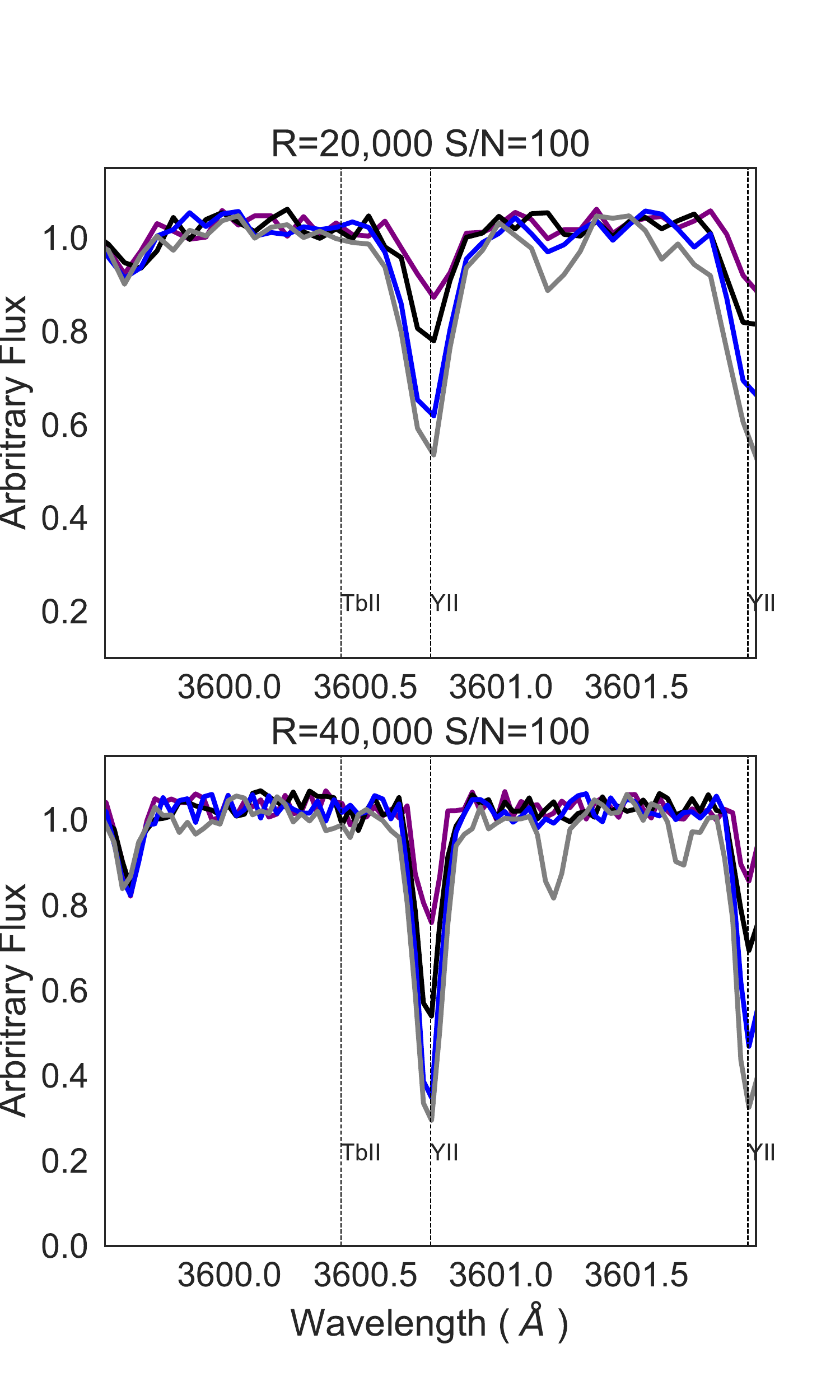}}
  \vspace{1.5mm}
    \caption{Simulated giant spectra ([Fe/H]\,$=$\,$-$3.0) for example diagnostic lines at $R$\,$=$\,20\,000 and 40\,000 (with S/N\,$=$\,100). Abundance variations of $-$0.5, 0.0, $+$0.5 and $+$1.0\,dex (in [X/H]) for the elements in Table~\ref{ablines} are shown in purple, black, blue and grey, respectively.}
    \label{giantlines}
\end{figure*}

\begin{table*}[h]
\caption{Summary of models used to investigate the diagnostic lines.}\label{models}
\centering                  
\begin{tabular}{lcc}
\hline\hline
Parameter                     & Dwarf   & Giant \\
\hline
T$_{\rm eff}$ [K]             & 5\,500  & 4\,500  \\ 
log\,$g$ [dex]                & 4.0     & 2.0   \\
$v_{\rm turb}$ [km\,s$^{-1}$] & \multicolumn{2}{c}{2.0} \\
\hline
{[Fe/H]}                      & \multicolumn{2}{c}{$-$3.0, $-$1.0} \\
$R$                           & \multicolumn{2}{c}{20\,000, 40\,000} \\
\hline\hline
& \\
\end{tabular}

\end{table*}
The models were convolved by a Gaussian with a full-width half maximum matched to that of the resolution at 3120\,\AA\ ($\Delta \lambda$\,$=$\,0.156\,\AA\ at $R$\,$=$\,20\,000, $\Delta \lambda$\,$=$\,0.078\,\AA\ at $R$\,$=$\,40\,000) and then binned to mimic the sampling by the detector, assuming 2.6 pixels per resolution element. To mimic real observations we introduced random noise in each of our models, to give simulated spectra with signal-to-noise (S/N) ratios of  50, 100, and 200 (per pixel). To illustrate the spectral richness of the near UV at the shortest wavelengths, an example section of one of the model spectra at $R$\,$=$\,20\,000 (prior to adding noise) is shown in Fig.~\ref{allspec1}. Examples of specific lines (Ge\,\1, Co\,\1, Ni\,\1, Y\,\2) in the simulated spectra of the giant star, varying $R$ at fixed S/N\,$=$\,100, are shown in Fig.~\ref{giantlines}.

\subsection{Results}
Each line in Table~\ref{ablines} was visually inspected for all four relative abundances (i.e. $-$0.5, 0.0, $+$0.5, $+$1.0) for both the dwarf and giant models at both resolving powers. For the purposes of the current study, we subjectively assessed the feasibility of obtaining an abundance estimate from each line on the basis of whether a `by-eye' fit to the simulated spectrum would enable constraints on the abundance (either a direct estimate or upper/lower limits). The example spectra in Fig.~\ref{giantlines} show some different situations where we can recover abundance information, e.g. from the strong and relatively isolated Ni\,\1 $\lambda$3598 line, to the richer region around Ge\,\1 $\lambda$3039.

Following inspection of each line, Figs~\ref{stellargrid} and \ref{stellargrid2} in the Appendix show which lines are feasible/useful, and Table~\ref{lines} summarises the number of lines available for each ion at $R$\,$=$\,20\,000. In terms of diagnostic lines commonly used in studies of neutron-capture elements, the only species that is not feasible is Ba\,\2, but there are stronger lines available at longer wavelengths that are more commonly used to estimate abundances (Ba\,\2 4554, 5853, 6496\,\AA) such that the near UV is not a necessity.

\begin{figure*}[h!]
    \centering
    \subfloat[Be\,{\scriptsize II}]{
  \includegraphics[trim= 0 0 0 2cm, clip, width=0.25\columnwidth]{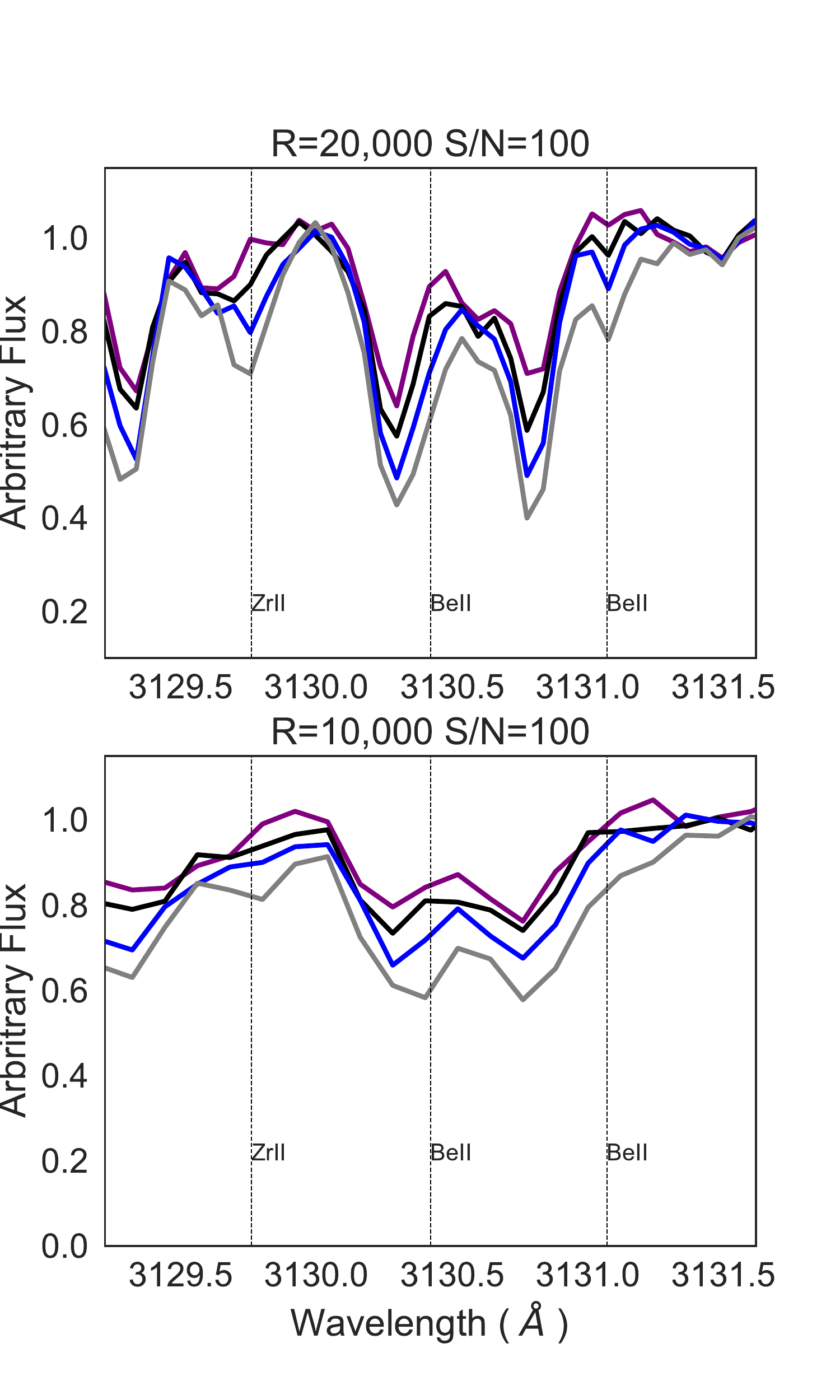}}
\subfloat[Co\,{\scriptsize I} $\lambda$3412]{
  \includegraphics[trim= 0 0 0 2cm, clip, width=0.25\columnwidth]{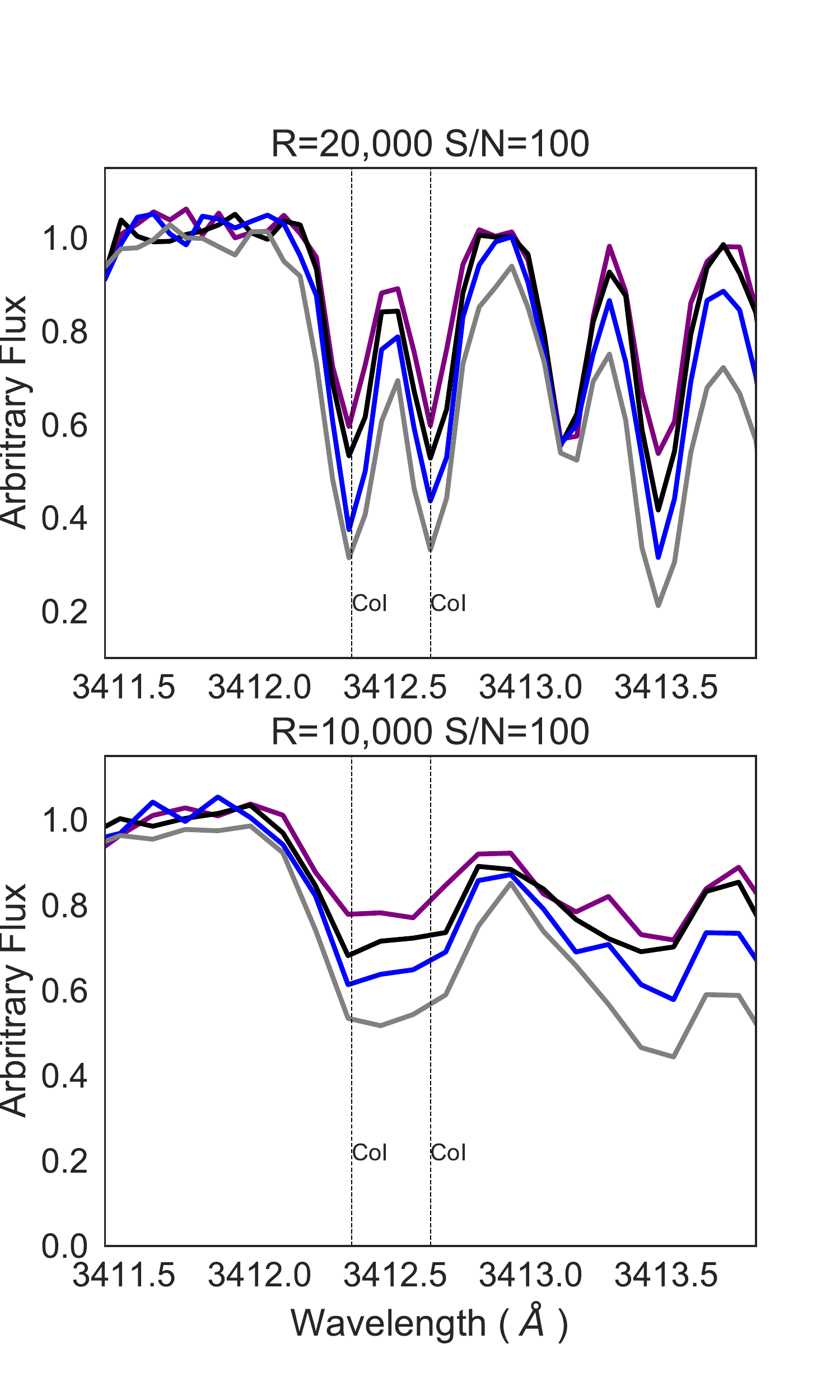}}
    \vspace{1.5mm}
    \caption{Illustrative comparison of two example pairs of diagnostic lines at $R$\,$=$\,10\,000 and 20\,000 (at S/N\,$=$\,100, for the [Fe/H]\,$=$\,$-$3.0 giant model). Abundance variations of $-$0.5, 0.0, $+$0.5 and $+$1.0\,dex (in [X/H]) for the elements in Table~\ref{ablines} are shown in purple, black, blue and grey, respectively.}
    \label{10k}
\end{figure*}

If the gains are sufficiently good in terms of sensitivity to work at $R$\,$=$\,20\,000 (cf. 40\,000), one might sensibly ask if even lower resolution observations are feasible.  We therefore also investigated simulated spectra with $R$\,$=$\,10\,000 to assess the impact of another factor of two in resolution. As demonstrated by the spectra in Fig.~\ref{10k}, this results in significant loss of information. Many close lines become strongly blended (e.g. the Co\,\1 doublet in the right-hand panel of the figure), and no constraint is possible on Be (left-hand panel). Indeed, at this lower resolution approximately 80\% of the lines in Table~\ref{ablines} are lost as useful abundance diagnostics, with only the most isolated and strongest lines remaining available. We therefore did not pursue such a low
resolution any further.

\subsection{Performance comparison}

As indicated by the results in Table~\ref{lines}, the majority of the diagnostic lines are accessible at $R$\,$=$\,20\,000 (provided there is sufficient S/N). Aside from optimisation of the instrument design, what the above comparisons neglect is the inherent loss in sensitivity of working at higher dispersion, i.e. a fairer comparison for the simulations shown in Fig.~\ref{giantlines} would be to show the resulting spectra for the same integration time.

The design philosophy for CUBES is to maximise the end-to-end efficiency of the instrument. 
It is unlikely that a similar efficiency could be obtained at the present time for a design at the higher resolving power. In short, $R$\,$=$\,20\,000 provides a combination of excellent sensitivity with sufficient resolution to undertake quantitative analysis of the large majority of the lines considered here.

 The faintest stars observed with VLT-UVES and Keck-HIRES to date for quantitative analysis in the near UV have $V$\,$\sim$\,12\,mag. For example, 2MASS~J18082002$-$5104378 ($V$\,$=$\,11.93, Schlaufman, Thompson \& Casey, 2018) was observed with ten 1\,h UVES exposures by Spite et al. (2019), giving S/N\,$\sim$\,70 near the Be lines. It is clear that going to fainter magnitudes (and to obtain better S/N) with UVES quickly starts to demand prohibitively long exposures/programmes of tens of hours per star. For comparison, using a developmental version of the CUBES Exposure Time Calculator (ETC)\footnote{\,http://archives.ia2.inaf.it/cubes/\#/etc}, observations of a metal-poor dwarf with $V$\,$=$\,16\,mag should provide a S/N\,$=$\,100 at 3130\,\AA\ in approximately 3$\times$1\,h exposures.

\section{Summary}

Near UV spectroscopy enables the study of a diverse range of elements for stellar astrophysics and of the chemical evolution of the Galaxy. Many of these are uniquely observable in the near UV, such as Be, Bi, and Os. Our study
of the elements accessible with observations at $R$\,$=$\,20\,000 compared to $R$\,$=$\,40\,000 reveals that nearly all species are feasible. In most instances the more dominant factor is S/N rather than resolving power, but the accuracy owing to blends may be reduced for some lines and requires more detailed simulations (e.g. C. Hansen et al. 2015). Reducing the resolution by a factor of two (to $R$\,$=$\,10\,000) would render most of the diagnostic lines unusable for abundance analysis. 

Informed by these results, the conceptual design of CUBES adopted $R$\,$\sim$\,20\,000 as its baseline. Quantifying the tolerances on this specification, including its variation with wavelength and performance for the light-element molecular features, is now underway as part of the Phase A study. From initial performance estimates, quantitative spectroscopy in the near-UV will be possible to at least three magnitudes deeper than current programmes (in the same exposure time). This will enable exciting new insights in our understanding of nucleosynthesis and the old stellar populations of the Milky Way.

\medskip
{\bf Acknowledgements:} We acknowledge support from the Global Challenges Research Fund (GCRF) from UK Research and Innovation (ST/R002630/1).
R.S. acknowledges support by the Polish National Science Centre through project 2018/31/B/ST9/01469. 

\bibliography{} 
\bibliographystyle{spiebib.bst} 
\noindent{\small \begin{tabular}{ll}
Abate et al. 2015, A\&A, 581, A62                    &  Lodders, 2003, ApJ, 591, 1220 \\                                         
Barbuy et al. 2014, Ap\&SS, 354, 191                 &  Matteucci et al. 2014, MNRAS, 438, 2177 \\                          
Barbuy, Trevisan \& de Almeida, 2018, PASA, 35, 46   &  Nishimura, Takiwaki \& Thielemann, 2015, ApJ, 810, 109 \\           
Beers \& Christlieb, 2005, ARA\&A, 43, 531           &  Nomoto, Kobayashi \& Tominaga, 2013, ARA\&A, 51, 457 \\             
Boesgaard et al. 1999, AJ, 117, 1549                 &  Pian et al. 2017, Nature, 551, 67  \\                               
Boesgaard et al. 2011, ApJ, 743, 140                 &  Placco et at al. 2015, ApJ, 812, 109 \\                             
Bovard et al. 2017, PhysRevD, 96, 124005             &  Plez, 2012, Astrophysics Source Code Library [ascl:1205.004] \\     
Busso, Gallino \& Wasserburg, 1999, ARA\&A, 37, 239  &  Primas et al. 2000a, A\&A, 362, 666 \\                              
Cayrel et al. 2001, Nature, 409, 691                 &  Primas et al. 2000b, A\&A, 364, L4 \\                               
Cescutti et al. 2015, A\&A, 577, A139                &  Ryabchikova et al. 2015, Physica Scripta, 90, 054005 \\            
Chiappini, 2013, AN, 334, 595                        &  Schlaufman, Thompson \& Casey, 2018, ApJ, 867, 98 \\		     
Choplin et al. 2016, A\&A, 593, A36                  &  Siegel et al. 2019, Nature, 569, 241 \\                            
Da Costa et al. 2019, MNRAS 489, 5900                &  Smartt et al. 2017, Nature, 551, 75 \\     			   
Dekker et al. 2000, Proc. SPIE, 4008, 534            &  Smiljanic et al. 2009, A\&A, 499, 103  \\  			    
Ernandes et al. 2018, A\&A, 616, A18                 &  Smiljanic et al. 2014, Ap\&SS, 354, 55   \\                        
Evans et al. 2016, Proc. SPIE, 9908, 9J              &  Sneden et al. 2003, ApJ, 591, 936 \\                                            
Evans et al. 2018, Proc. SPIE, 10702, 2E             &  Spite et al. 2019, A\&A, 624, A44 \\
Frischknecht et al. 2012, A\&A, 538, 2               &  Starkenburg et al. 2017, MNRAS, 471, 2587  \\ 
Gustafsson et al. 2008, A\&A, 486, 951               &  Tominaga, Iwamoto \& Nomoto, 2014, ApJ, 785, 98  \\            
Hansen, C. et al. 2015, AN, 336, 665                 &  Umeda \& Nomoto, 2002, ApJ, 565, 385 \\                 
Hansen, T. et al. 2015, ApJ, 807, 173                &  Umeda \& Nomoto, 2005, ApJ, 619, 427  \\                             
Hansen et al. 2019, A\&A, 623, 128                   &  Watson et al. 2019, Nature, 574, 497 \\                            
Hill et al. 2002, A\&A, 387, 560                     &  Winteler et al. 2012, ApJ, 750, L22 \\                            
Limongi \& Chieffi, 2003, ApJ, 592, 404              &  Wolf et al. 2018, PASA, 35, 10 \\    
Limongi \& Chieffi, 2018, ApJ, 237, 13               &  Woosley, Heger \& Weaver, 2002, RvMP, 74, 1015 \\                   
\end{tabular}
}

\newpage
\appendix    

\section{Supporting material}




\begin{table*}[h]
\caption{ List of near-UV absorption lines considered.}   
\label{ablines} 
\centering                  
\begin{tabular}{ l  c | l  c | l c } 
\hline\hline             
  Ion   &  Wavelength ({\rm \AA}) &   Ion   &  Wavelength ({\rm \AA}) &   Ion   &  Wavelength ({\rm \AA})   \\
\hline  
Be\,\2 & 3130.42   & Zr\,\2 & 3357.26  & Nd\,\2 & 3826.41  \\ 
Be\,\2 & 3131.07   & Zr\,\2 & 3404.83  & Nd\,\2 & 3838.98  \\
Sc\,\2 & 3576.34   & Zr\,\2 & 3408.08  & Sm\,\2 & 3568.27  \\
Sc\,\2 & 3590.47   & Zr\,\2 & 3410.24  & Sm\,\2 & 3796.75  \\
Ti\,\1 & 3998.64   & Zr\,\2 & 3430.53  & Sm\,\2 & 3896.97  \\
Ti\,\2 & 3321.70   & Zr\,\2 & 3438.23  & Eu\,\2 & 3724.93  \\
Ti\,\2 & 3343.76   & Zr\,\2 & 3457.56  & Eu\,\2 & 3819.67  \\ 
Ti\,\2 & 3491.05   & Zr\,\2 & 3458.93  & Eu\,\2 & 3907.11  \\ 
V\,\2  & 3951.96   & Zr\,\2 & 3479.02  & Eu\,\2 & 3930.40  \\
Cr\,\1 & 3578.68   & Zr\,\2 & 3479.39  & Gd\,\2 & 3549.36  \\                     
Mn\,\2 & 3441.99   & Zr\,\2 & 3481.15  & Gd\,\2 & 3557.06  \\
Mn\,\2 & 3460.32   & Zr\,\2 & 3496.20  & Gd\,\2 & 3712.70  \\
Mn\,\2 & 3482.90   & Zr\,\2 & 3505.67  & Gd\,\2 & 3768.40  \\
Mn\,\2 & 3488.68   & Zr\,\2 & 3506.05  & Tb\,\2 & 3600.41  \\
Mn\,\2 & 3495.83   & Zr\,\2 & 3525.81  & Tb\,\2 & 3702.85  \\
Mn\,\2 & 3497.53   & Zr\,\2 & 3549.51  & Dy\,\2 & 3531.71  \\
Co\,\1 & 3412.34   & Zr\,\2 & 3551.95  & Dy\,\2 & 3536.02  \\
Co\,\1 & 3412.63   & Zr\,\2 & 3556.59  & Dy\,\2 & 3550.22  \\
Co\,\1 & 3449.16   & Zr\,\2 & 3576.85  & Dy\,\2 & 3563.15  \\
Co\,\1 & 3529.03   & Zr\,\2 & 3588.31  & Dy\,\2 & 3694.81  \\ 
Co\,\1 & 3842.05   & Zr\,\2 & 3607.37  & Dy\,\2 & 3757.37  \\
Co\,\1 & 3845.47   & Zr\,\2 & 3614.76  & Dy\,\2 & 3944.68  \\
Ni\,\1 & 3437.28   & Zr\,\2 & 3751.59  & Dy\,\2 & 3996.69  \\
Ni\,\1 & 3483.77   & Zr\,\2 & 3766.82  & Ho\,\2 & 3466.01  \\
Ni\,\1 & 3500.85   & Zr\,\2 & 3836.76  & Ho\,\2 & 3796.80  \\
Ni\,\1 & 3597.71   & Zr\,\2 & 3998.96  & Ho\,\2 & 3890.65  \\
Ni\,\1 & 3807.14   & Nb\,\2 & 3028.44  & Er\,\2 & 3692.65  \\
Cu\,\1 & 3247.53   & Nb\,\2 & 3215.59  & Er\,\2 & 3729.52  \\
Cu\,\1 & 3273.95   & Nb\,\2 & 3225.47  & Er\,\2 & 3786.84  \\
Zn\,\1 & 3075.90   & Mo\,\1 & 3864.10  & Er\,\2 & 3830.48  \\ 
Zn\,\1 & 3302.58   & Ru\,\1 & 3436.74  & Er\,\2 & 3896.23  \\
Zn\,\1 & 3345.01   & Ru\,\1 & 3498.94  & Er\,\2 & 3906.31  \\ 
Ge\,\1 & 3039.07   & Ru\,\1 & 3742.28  & Tm\,\2 & 3701.36  \\ 
Y\,\2  & 3549.01   & Ru\,\1 & 3798.90  & Tm\,\2 & 3795.76  \\ 
Y\,\2  & 3584.52   & Ru\,\1 & 3799.35  & Tm\,\2 & 3848.02  \\
Y\,\2  & 3600.74   & Rh\,\1 & 3396.82  & Yb\,\2 & 3694.20  \\ 
Y\,\2  & 3601.91   & Rh\,\1 & 3434.89  & Hf\,\2 & 3276.85  \\
Y\,\2  & 3611.04   & Rh\,\1 & 3692.36  & Hf\,\2 & 3399.79  \\ 
Y\,\2  & 3774.33   & Rh\,\1 & 3700.91  & Hf\,\2 & 3719.28  \\ 
Y\,\2  & 3788.69   & Pd\,\1 & 3242.70  & Os\,\1 & 3058.66  \\ 
Y\,\2  & 3818.34   & Pd\,\1 & 3404.58  & Ir\,\1 & 3220.78  \\  
Y\,\2  & 3950.35   & Pd\,\1 & 3516.94  & Ir\,\1 & 3800.12  \\
Zr\,\2 & 3054.84   & Ag\,\1 & 3280.68  & Pb\,\1 & 3683.46  \\
Zr\,\2 & 3095.07   & Ag\,\1 & 3382.90  & Bi\,\1 & 3024.64  \\
Zr\,\2 & 3125.92   & Sn\,\1 & 3801.01  & Th\,\2 & 3351.23  \\
Zr\,\2 & 3129.76   & Ba\,\2 & 3891.78  & Th\,\2 & 3433.99  \\
Zr\,\2 & 3273.05   & La\,\2 & 3794.77  & Th\,\2 & 3435.98  \\
Zr\,\2 & 3279.26   & La\,\2 & 3949.10  & Th\,\2 & 3469.92  \\
Zr\,\2 & 3284.71   & La\,\2 & 3988.51  & Th\,\2 & 3539.59  \\
Zr\,\2 & 3305.15   & La\,\2 & 3995.74  & Th\,\2 & 3675.57  \\
Zr\,\2 & 3334.62   & Ce\,\2 & 3999.24  & U\,\2  & 3859.57  \\
Zr\,\2 & 3344.79   & Nd\,\2 & 3784.24  &    &     \\
Zr\,\2 & 3356.09   & Nd\,\2 & 3810.48  &    &     \\
\hline
\hline                          
\end{tabular}
\end{table*}

\begin{table*}[h]
\caption{ Number of our selected lines that are detectable for each ion in the CUBES wavelength range for both a G-type dwarf and a K-type giant with [Fe/H]\,$=$\,$-$3.0 and [Fe/H]\,$=$\,$-$1.0 with $R$\,$\sim$\,20\,000. Those flagged with an asterisk
are ions for which no lines are feasible for quantitative analysis in our simulated
spectra.\\}   
\label{lines} 
\centering                  
\begin{tabular}{ l | c| c || c | c } 
\hline\hline
& \multicolumn{2}{c||}{~~[Fe/H]\,$=$\,$-$3.0~~} &\multicolumn{2}{c}{~~[Fe/H]\,$=$\,$-$1.0~~}
\\
\hline
Ions~~~~~~  &   Giant   &  Dwarf    &   Giant   &  Dwarf    \\
\hline \hline
Be\,\2     &   2/2       &     2/2     &   2/2       &     2/2    \\
Sc\,\2     &   2/2       &     2/2     &   2/2       &     2/2    \\
Ti\,\1     &   1/1       &     1/1     &   1/1       &     1/1 \\
Ti\,\2     &   3/3       &     3/3     &   3/3       &     3/3 \\
V\,\2      &   1/1       &     1/1     &   1/1       &     1/1 \\
Cr\,\1     &   1/1       &     1/1     &   0/1       &     1/1 \\
Mn\,\2     &   4/6       &     5/6     &   4/6       &     4/6 \\
Co\,\1     &   6/6       &     6/6     &   6/6       &     6/6 \\
Ni\,\1     &   5/5       &     5/5     &   5/5       &     4/5 \\
Cu\,\1     &   2/2       &     2/2     &   0/2       &     2/2 \\
Zn\,\1     &   1/3       &     3/3     &   1/3       &     3/3 \\
Ge\,\1     &   1/1       &     1/1     &   1/1       &     1/1 \\
Y\,\2      &   3/6       &     4/6     &   4/6       &     4/6 \\
Zr\,\2     &   ~29/34~   &     ~30/34~ &   ~12/34~   &     ~19/34~ \\
Nb\,\2     &   3/3       &     1/3     &   2/3       &     2/3 \\
Mo\,\1     &   1/1       &     0/1     &   1/1       &     1/1 \\
Ru\,\1     &   3/5       &     0/5     &   3/5       &     3/5 \\
Rh\,\1     &   2/4       &     0/4     &   1/4       &     1/4 \\
Pd\,\1     &   3/3       &     0/3     &   2/3       &     2/3 \\
Ag\,\1     &   2/2       &     0/2     &   2/2       &     0/2 \\
Sn\,\1     &   1/1       &     0/1     &   0/1       &     0/1 \\
Ba\,\2*    &   0/1       &     0/1     &   0/1       &     0/1 \\
La\,\2     &   3/4       &     2/4     &   1/4       &     2/4 \\
Ce\,\2     &   1/1       &     1/1     &   1/1       &     1/1 \\
Nd\,\2     &   4/4       &     0/4     &   3/4       &     4/4 \\
Sm\,\2*    &   0/3       &     0/3     &   0/3       &     0/3 \\
Eu\,\2     &   1/3       &     1/3     &   1/3       &     2/3 \\
Gd\,\2     &   0/4       &     0/4     &   1/4       &     0/4 \\
Tb\,\2     &   2/2       &     0/2     &   1/2       &     2/2 \\
Dy\,\2     &   8/8       &     0/8     &   6/8       &     5/8 \\
Ho\,\2*    &   0/3       &     0/3     &   0/3       &     0/3 \\
Er\,\2     &   3/6       &     0/6     &   2/6       &     2/6 \\
Tm\,\2     &   0/3       &     0/3     &   0/3       &     0/3 \\
Yb\,\2*    &   0/1       &     0/1     &   0/1       &     0/1 \\
Hf\,\2     &   1/3       &     0/3     &   1/3       &     1/3 \\
Os\,\1     &   1/1       &     0/1     &   1/1       &     1/1 \\
Ir\,\1     &   2/2       &     0/2     &   1/2       &     2/2 \\
Pb\,\1     &   0/1       &     0/1     &   0/1       &     1/1 \\
Bi\,\1     &   0/1       &     0/1     &   1/1       &     1/1 \\
Th\,\2     &   0/6       &     0/6     &   1/6       &     1/6 \\
U\,\2      &   1/1       &     1/1     &   0/1       &     1/1 \\
\hline
\hline                          
\end{tabular}
\end{table*}

\begin{figure}[!h]
    \centering
    \subfloat[Lines from Be\,{\scriptsize II} to Zr\,{\scriptsize II}.]{
  \includegraphics[trim= 0 0 1.7cm 0, clip, width=8.0cm]{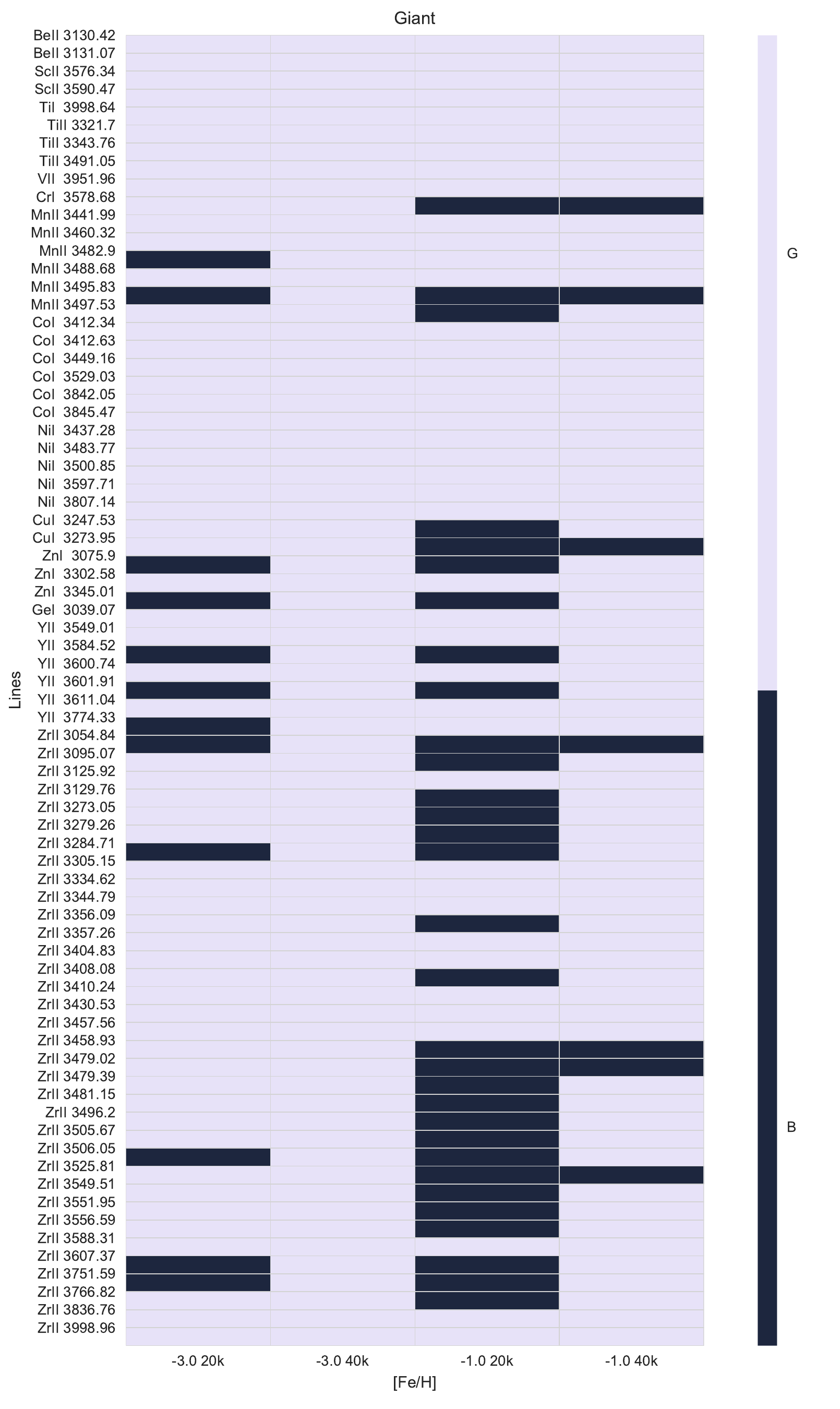}
}
\subfloat[Lines from Nb\,{\scriptsize II} to U\,{\scriptsize II}.]{
  \includegraphics[trim= 0 0 1.7cm 0, clip, width=8.0cm]{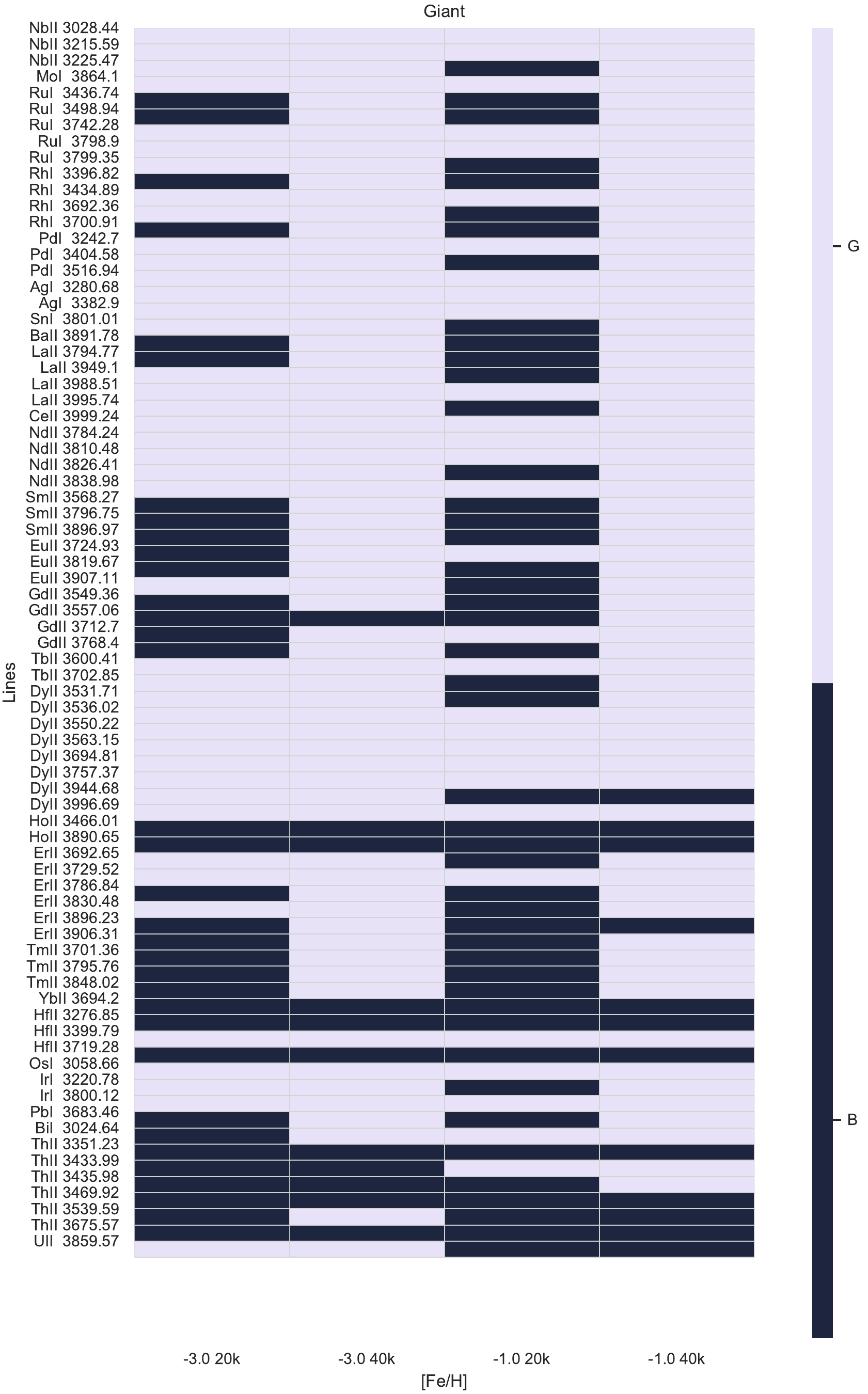}
}
\smallskip
    \caption{Detectable lines in the simulated K-type giant spectra with [Fe/H]\,$=$\,$-$3.0 and $-$1.0 for $R$\,$=$\,20,000 and 40,000. Darker shaded boxes indicate where it is not possible to discern differences between abundance variations of $-$0.5, 0.0, $+$0.5 and $+$1.0\,dex (in [X/H]).}
    \label{stellargrid}
\end{figure}

\begin{figure}[!h]
    \centering
    \subfloat[Lines from Be\,{\scriptsize II} to Zr\,{\scriptsize II}.]{
  \includegraphics[trim= 0 0 1.7cm 0, clip, width=8.0cm]{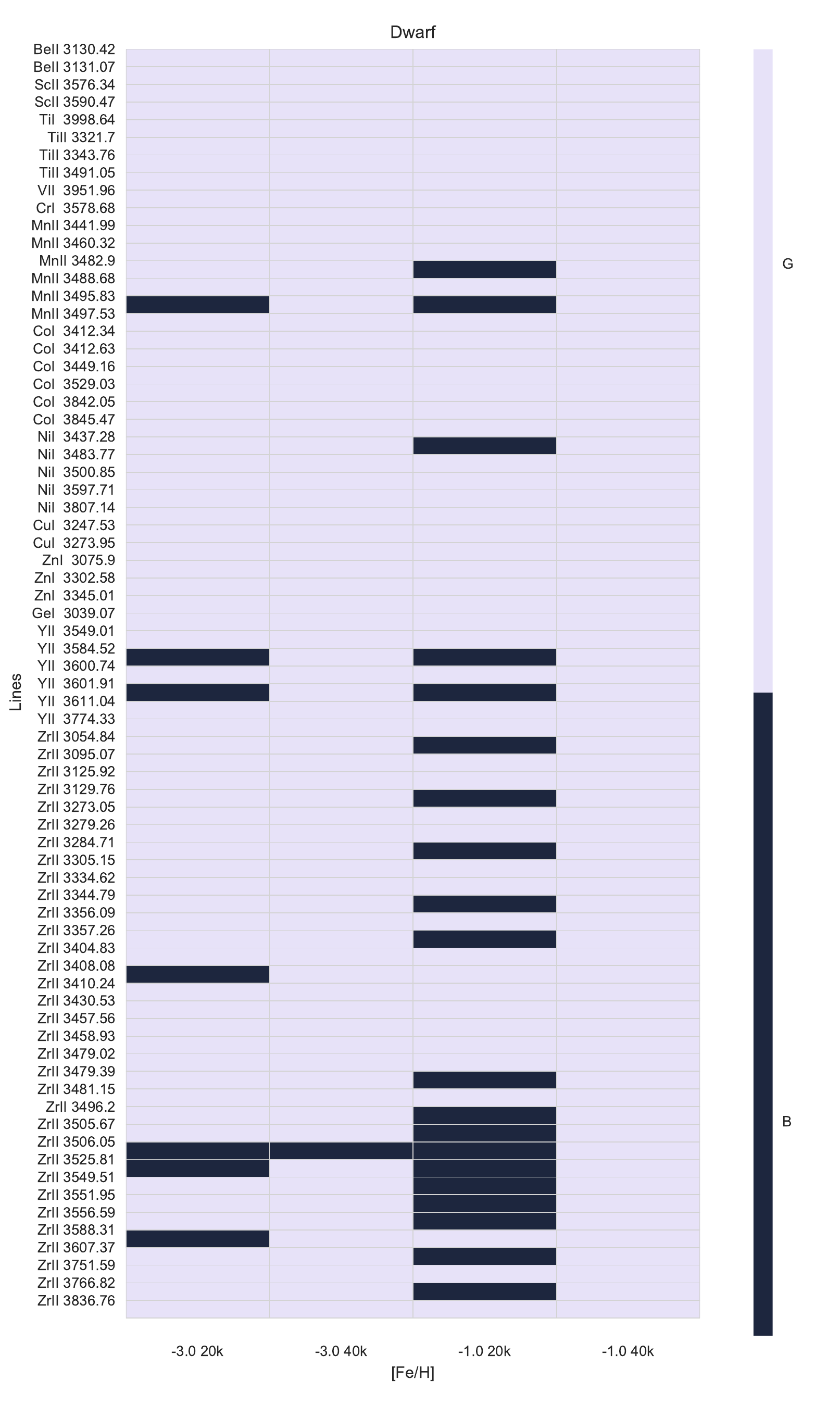}
}
\subfloat[Lines from Nb\,{\scriptsize II} to U\,{\scriptsize II}.]{
  \includegraphics[trim= 0 0 1.7cm 0, clip, width=8.0cm]{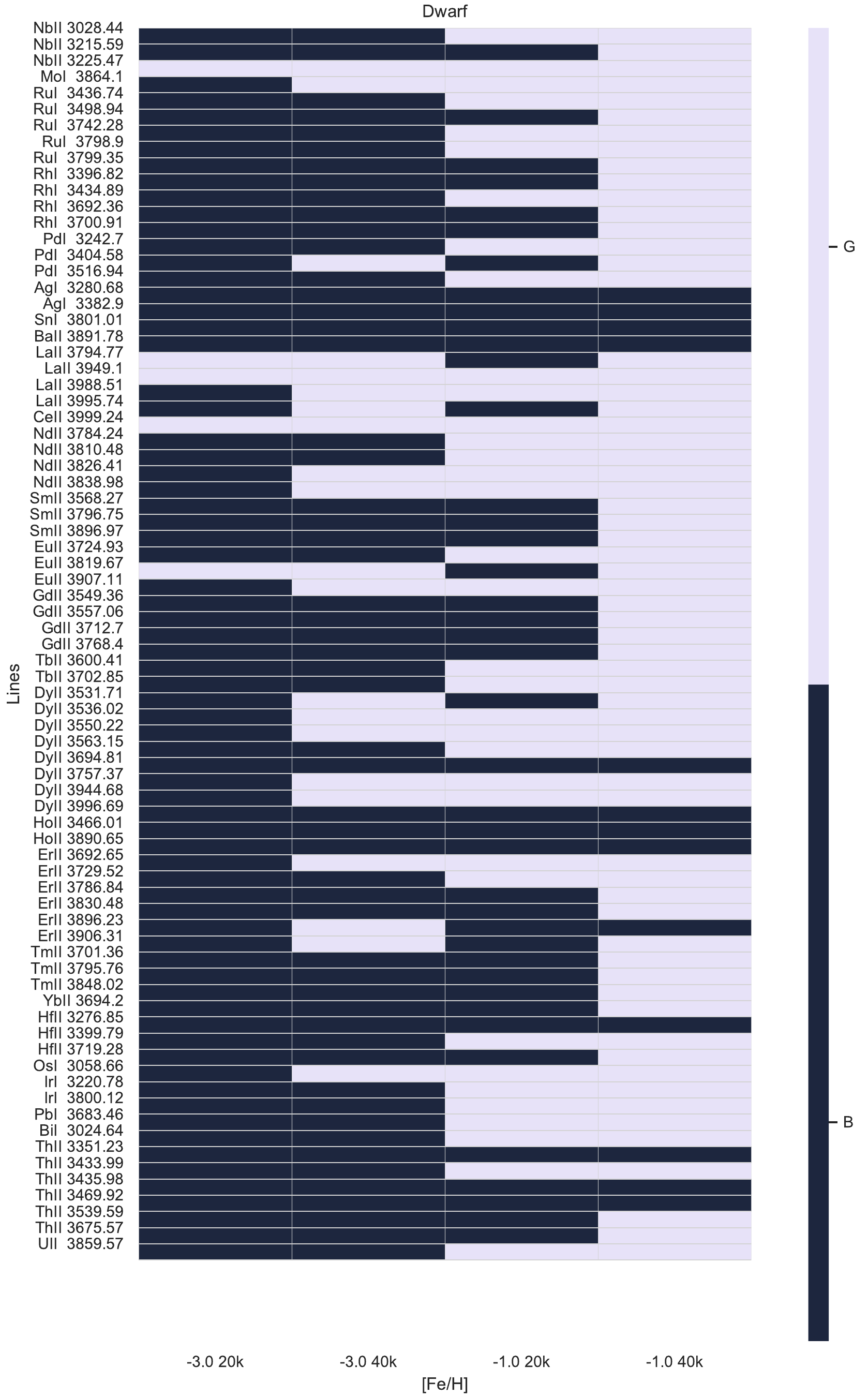}
}
\smallskip
    \caption{Detectable lines in the simulated G-type dwarf spectra with [Fe/H]\,$=$\,$-$3.0 and $-$1.0 for $R$\,$=$\,20,000 and 40,000. Darker shaded boxes indicate where it is not possible to discern differences between abundance variations of $-$0.5, 0.0, $+$0.5 and $+$1.0\,dex (in [X/H]).}
    \label{stellargrid2}
\end{figure}

\end{document}